\documentclass[
preprint
]{elsarticle}
\journal{Soft Matter}
\usepackage{graphicx}
\usepackage{amsmath}
\usepackage{epstopdf}
\usepackage[colorlinks=false]{hyperref}
\usepackage{color}
\usepackage{lineno}

\newcommand{\dcdr}{(\partial c/\partial r)_R}
\newcommand{\DCDR}{\left(\frac{\partial c}{\partial r}\right)_R}

\begin{document}

\title{Gas bubble dynamics in soft materials}

\author[uoguelph]{J. M. Solano-Altamirano\corref{cor1}}
\ead{jmsolanoalt@gmail.com}

\author[uoguelph]{John D. Malcolm}
\ead{malcolmj@uoguelph.ca}

\author[uoguelph]{Saul Goldman}
\ead{sgoldman@uoguelph.ca}

\address[uoguelph]{Dept. of Chemistry, the Guelph-Waterloo Centre for Graduate Work in Chemistry and the Guelph-Waterloo Physics Institute, University of Guelph, Guelph, Ontario, N1G 2W1, Canada. Phone: +1 5198244120 Ext. 53830, Fax: +1 5197661499}

\cortext[cor1]{Corresponding author.}

\date{\today}                                           

\begin{abstract}
Epstein and Plesset's seminal work on the rate of gas bubble dissolution and growth in a simple liquid is generalized to render it applicable to a gas bubble embedded in a soft elastic medium. Both the underlying diffusion equation and the expression for the gas bubble pressure were modified to allow for the non-zero shear modulus of the elastic medium. The extension of the diffusion equation results in a trivial shift (by an additive constant) in the value of the diffusion coefficient, and does not change the form of the rate equations. But the use of a Generalized Young-Laplace equation for the bubble pressure resulted in significant differences on the dynamics of bubble dissolution and growth, relative to a simple liquid medium. Depending on whether the salient parameters (solute concentration, initial bubble radius, surface tension, and shear modulus) lead to bubble growth or dissolution, the effect of allowing for a non-zero shear modulus in the Generalized Young-Laplace equation is to speed up the rate of bubble growth, or to reduce the rate of bubble dissolution, respectively. The relation to previous work on visco-elastic materials is discussed, as is the connection of this work to the problem of Decompression Sickness (specifically, ``the bends''). Examples of tissues to which our expressions can be applied are provided. Also, a new phenomenon is predicted whereby, for some parameter values, a bubble can be metastable and persist for long times, or it may grow, when embedded in a homogeneous under-saturated soft elastic medium.
\end{abstract}



\maketitle

\section{Introduction}

More than sixty years ago, Epstein and Plesset published a seminal article in which they showed how to semi-quantitatively estimate the rate of gas bubble growth or dissolution, for a bubble embedded in a liquid medium containing the dissolved gas of which the bubble is comprised \cite{bib:epspless1950}. They applied their expressions to air bubbles suspended in water, containing dissolved air. Their rate expressions have been experimentally found to be largely correct, and the precise degree of validity of their model remains the subject of active research \cite{bib:duncan2004,bib:duncan2006,bib:enriquez2013}. For small dissolving bubbles, the predictions of Epstein and Plesset's model were found to be within about 9\% of the observed values for the surface tension and saturation level dependencies \cite{bib:duncan2004} when the surrounding medium is a simple liquid.

Their work was subsequently applied to a variety of problems that arise in volcanology \cite{bib:watson1982,bib:kuchma2009,bib:gorkuchma2009}, cavitation in liquids \cite{bib:plessprosp1977,bib:neppiras1979} and physiology \cite{bib:srinivasan1999,bib:jmsolano2014}. Here we extend Epstein and Plesset's approach to a gas bubble embedded in a soft slightly compressible elastic material. This extension is required in order to correctly model gas bubble growth and dissolution in molten magma, and in soft extravascular tissue in the human body. The latter application arises in the problem of Decompression Sickness, which is of interest to us \cite{bib:jmsolano2014,bib:goldman2007}. 

Decompression Sickness arises due to the growth of gas bubbles in blood and tissues, as a consequence of an overly rapid decompression (\textit{i.e.}, drop in external pressure), which may arise from an overly rapid ascent from a scuba dive, or from too rapid a drop in external pressure in aviation or space exploration. Two basic causative mechanisms of Decompression Sickness are currently distinguished, depending on whether the expanding bubbles are in arterial circulation, or whether they are lodged in extravascular tissue \cite{bib:francismitchell2003}. The expansion of gas bubbles that get into arterial circulation --- Arterial Gas Emboli (AGEs) --- is believed to initiate Cerebral, Spinal, Inner Ear, and Skin Decompression sickness, while the expansion of extravascular (or ``autochthonous'') bubbles is believed to be responsible for joint and musculoskeletal pain (colloquially, ``the bends'') \cite{bib:francismitchell2003}. In an earlier article \cite{bib:jmsolano2014} we applied Epstein and Plesset's work to AGEs in relation to their connection to Inner Ear Decompression Sickness. We would like to extend our work to include the dynamics (growth and dissolution) of gas bubbles lodged in soft extravascular tissue. Consequently, we focus here on developing the theoretical tools needed to do this. These bubbles are lodged in a medium which much more closely resembles soft elastic matter, than it does a simple liquid (one without shear resistance or intrinsic shape). While Epstein and Plesset's work was essentially directly applicable to AGEs (since arterial blood is a liquid medium much like water), it must be significantly modified before it can be applied to a gas bubble lodged in soft tissue.

We derive generalized rate equations that take into account the influence of a non-zero shear modulus in the medium. This manifests itself both on the magnitude of the internal gas bubble pressure and on the diffusion equation used for the medium. We illustrate our expressions by using them to predict the dynamics of growth and dissolution of an embedded gas bubble in a soft material with properties similar to that of many soft tissues in the human body. We also briefly compare our work with earlier work on gas bubbles in various types of visco-elastic media.

\section{Theory}

Epstein and Plesset's rate laws for gas bubble growth and dissolution are obtained from equations (\ref{eq:yleq})-(\ref{eq:henryslawarb}) and (\ref{eq:diffeqor}), given below. Combining Eqs. (\ref{eq:yleq})-(\ref{eq:henryslawarb}) provides the relation between $dR/dt$ and $\dcdr$, and the solution of Eq. (\ref{eq:diffeqor}) provides the expression(s) for $\dcdr$.

The Young-Laplace equation for the gas pressure inside a bubble embedded in a simple liquid (\textit{i.e.} one without shear forces) is:
\begin{equation}\label{eq:yleq}
   P_B=P_e+\frac{2\gamma}{R},
\end{equation}
where $P_B$ is the gas pressure inside the bubble, $P_e$ is the exterior pressure that acts on the medium, $\gamma$ the surface tension at the bubble-medium interface, and $R$ is the radius of the bubble. 

Fick's law for the rate of solute transfer across a spherical interface is given by:
\begin{equation}\label{eq:fickslaw}
   \frac{dn}{dt}=4\pi DR^2\DCDR,
\end{equation}
where $n$ is the number of moles of gas in the bubble, $t$ the time, $D$ is the diffusion coefficient of the solute in the medium, and $\dcdr$ is the dissolved solute concentration gradient at the surface of the bubble.

\begin{figure}[ht!]
\centering
\includegraphics[width=0.45\textwidth]{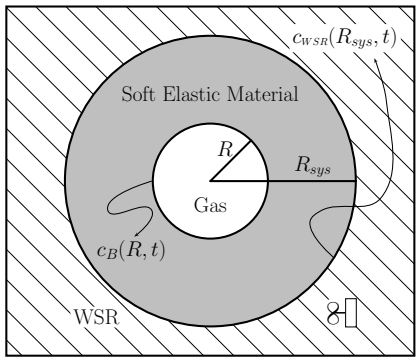}
\caption{A gas bubble of radius $R$ surrounded by an elastic medium through which dissolved gas diffuses either toward or away from the bubble. The elastic medium is split into an inner spherical shell whose thickness is ($R_{sys}-R$), and a ``well stirred region'' (WSR), wherein the dissolved solute concentration ($c_{WSR}$) is constant and uniform. The scheme represents a ``snapshot'' of the dynamical system, which is assumed to always be in a quasi-static state, throughout the growth or dissolution of the bubble.}\label{fig:physschem}
\end{figure}

Henry's law is assumed to apply at all the boundaries of the system:
\begin{equation}\label{eq:henryslawarb}
   c(R^*,t)=\frac{P(R^*,t)}{K_H}.
\end{equation}
Here $R^*$ represents the distance from the center of the system to any of the boundaries (see Fig. \ref{fig:physschem}), $c(R^*,t)$ is the dissolved gas concentration at $(R^*,t)$ (in units mol/l), $K_H$ the Henry's constant for the gas dissolved in the medium and contained in the bubble, and $P(R^*,t)$ is the dissolved gas partial pressure at the boundary whose distance is $R^*$ from the system center. $K_H$ is an equilibrium constant that is related to the solubility of the gas in the medium \cite{bib:hildebrand1964}. It gives the ratio of the gas partial pressure to its concentration in solution at equilibrium.

To derive an expression(s) for $\dcdr$, we start by considering the full diffusion equation for a two-component fluid in the absence of elastic effects \cite{bib:landau1959}:
\begin{equation}\label{eq:fulldiffeq}
   \frac{\partial c}{\partial t}=-\nabla\cdot\boldsymbol{J}-\nabla\cdot\left(c\boldsymbol{v}\right).
\end{equation}

Here $\boldsymbol{J}$ is the flux of dissolved material within a differential volume element of the elastic medium, and $\boldsymbol{v}$ is the velocity of the volume element relative to the bubble. In Eq. (\ref{eq:fulldiffeq}) it is assumed that any change in density stemming from a change of concentration of the dissolved substance can be neglected. The second term on the right of Eq. (\ref{eq:fulldiffeq}) provides the contribution to $\partial c/\partial t$ due to any motion of the volume element (\textit{e.g.} convective motion due to mixing or flow) relative to the bubble.

In order to derive analytic expressions for the rate of bubble growth/dissolution, Epstein and Plesset made two fundamental assumptions. One was to omit the contribution due to relative motion of the medium \textit{i.e.}, they took $\boldsymbol{v}$ in Eq. (\ref{eq:fulldiffeq}) to be zero. The other involved a separation of time scales. This arose through their (tacit) assumption that any perturbation in the system --- specifically, the transfer of a small amount of solute across the bubble interface --- is followed by an instantaneous re-equilibration of the solute distribution in the entire system --- \textit{i.e.} both in the bubble and in the medium. Thus, equations (\ref{eq:yleq})-(\ref{eq:henryslawarb}) are taken apply at all times. We will here refer to their second approximation as the ``quasi-static approximation'', since it approximates the bubble as static (or growing infinitesimally slowly), relative to the very rapid re-distribution rate of the solute within the bubble and the surrounding medium. As mentioned above, the overall error that results from both of these approximations, made simultaneously, is within about 9\% for the predicted dissolutions times for small bubbles in an under-saturated medium \cite{bib:duncan2004}. Bubble growth in slightly super-saturated solutions also appears to be driven mostly by diffusive processes \cite{bib:enriquez2013}. Hence, the quasi-static approximation is in fairly good agreement with experimental observations both for slightly super-saturated and slightly under-saturated solutions. Good reviews that critically analyze the physical conditions under which the approximations made by Epstein and Plesset are viable and/or useful is provided in Refs. \citenum{bib:kuchma2009} and \citenum{bib:gorkuchma2009}.

With these approximations, the full diffusion equation (\ref{eq:fulldiffeq}) is reduced to the familiar diffusion equation:
\begin{equation}\label{eq:diffeqor}
   \frac{\partial c}{\partial t}=D\nabla^2c.
\end{equation}
In this work, we also make both of these assumptions, for the same reasons as in Epstein and Plesset's work. In addition, neglect of the medium's motion is further necessitated by our application of linear elasticity theory, which requires that our system undergoes only coherent deformations. By a ``coherent deformation'' is meant that the coordinates of the deformed body are isomorphic functions of the coordinates of the un-deformed body \cite{bib:sokolnikov1956}.

In order to allow for the effects of a non-zero shear modulus in the medium we need, in addition to the above approximations, extensions of Eqs. (\ref{eq:yleq}) and (\ref{eq:diffeqor}). The extension of Eq. (\ref{eq:diffeqor}) for describing the diffusion of a gas within a solid has been known for some time (see Refs. \citenum{bib:aifantis1978} and references therein):
\begin{subequations}\label{eq:aifantisdiffeq}
   \begin{eqnarray}
      \frac{\partial c}{\partial t}&=&D^*\nabla^2c-M\nabla\sigma\cdot\nabla c.\label{eq:dcdtdast}\\
      D^*&=&D+N\sigma,
   \end{eqnarray}
\end{subequations}
In Eqs. (\ref{eq:aifantisdiffeq}), $M$ and $N$ are phenomenological constants characteristic of the medium, $\sigma$ is the trace of the stress tensor of the medium, and $D^*$ is an effective diffusion constant. However, for the isotropic elastic media which are considered here, these equations simplify. Since the trace of the stress tensor in an isotropic medium of arbitrary shape is a constant \cite{bib:timoshenko1951,bib:landau1989}, the second term on the right-hand side of Eq. (\ref{eq:dcdtdast}) vanishes for isotropic media. The resultant equation is:
\begin{equation}\label{eq:diffeqdast}
   \frac{\partial c}{\partial t}=D^*\nabla^2c.
\end{equation}

In what follows, we provide the extension of Eq. (\ref{eq:yleq}), and subsequently use it derive a generalization of Epstein and Plesset's solution for a soft elastic medium.

We consider a gas bubble embedded in an elastic medium, as illustrated in Fig. \ref{fig:physschem}.This  physical model is identical to the one introduced by Epstein and Plesset, except that we here allow for a positive shear resistance in the diffusive medium. The general purpose of the approximate model shown in Fig. \ref{fig:physschem} is to simplify the problem by physically separating the regions wherein diffusion and convection (or mixing) are presumed to occur. Diffusion alone is presumed to occur in the diffusive region via a concentration gradient within this region. The well-stirred region, which is comprised of the same elastic material as that of the diffusive region, is presumed to be perfectly well-mixed, \textit{i.e.} it has no solute concentration gradient(s).

The entire system is taken to be in a quasi-static state at all times. Dissolved gas diffuses either toward or away from the bubble due to a solute concentration gradient within the diffusion shell that surrounds the bubble. This gradient stems from the solute concentrations at $R_{sys}$ and at $R$ being different. The concentration is fixed at the constant value $c_{WSR}$ at $R_{sys}$, and it is determined by the bubble pressure $P_B$, and Henry's law, at $R$.  Since $P_B$ is a function of $R$ (see Eqs. (\ref{eq:gylpbfull}), (\ref{eq:gyleq3r}), and (\ref{eq:gyleq2r})), and $R$ varies with $t$, $c_B$ will vary as the bubble shrinks or grows. The bubble will dissolve if $c_B > c_{WSR}$, it will grow if $c_B < c_{WSR}$, and it will be stable or meta-stable, and persist for relatively long times (or indefinitely) if $c_B=c_{WSR}$.

\subsection{Modification of the diffusion equation to allow for elastic effects}

Recently, Goldman generalized the Young-Laplace equation by considering the effect of an elastic body on the internal pressure of a gas bubble embedded within it \cite{bib:goldman2009}. For a spherical elastic shell containing a spherical embedded gas bubble at its center, the result was found to be given by the system of coupled equations:
\begin{subequations}
\label{eq:gpbcoupled}
\begin{equation}\label{eq:gylpbfull}
   P_B=P_eh(\nu)+4a_1G\left[1-\nu h(\nu)\right]+\frac{2\gamma}{R},
\end{equation}
\begin{equation}
   h(\nu)=\frac{1+\delta}{1+\nu\delta},\qquad\delta=\frac{4G}{3K},
\end{equation}
\begin{equation}
   \nu=\left(\frac{R}{R_{sys}}\right)^3,
\end{equation}
\begin{equation}
   R_{sys}^3=R^3+(3V_{in}^{(el)}/4\pi)\left(1+3a_2\right),
\end{equation}
\begin{equation}
   a_2=\frac{4Ga_1\nu-P_e}{3K+4G\nu}.
\end{equation}
\end{subequations}
Here, $V_{in}^{(el)}$ is the initial volume of the elastic shell (\textit{i.e.} prior to compression by the application of non-zero pressures to its surfaces), $G$ is the shear modulus of the elastic medium, $K$ is its modulus of compression (\textit{aka} bulk modulus), $P_e$ (as in Eq. (\ref{eq:yleq})) is the external pressure applied at the outer radius of the spherical shell, and $a_1$ is a constant related to the volumetric change of the gas in the bubble due to compression. For ideal gases, $a_1=-1/3$ \cite{bib:goldman2009}.

For reasons given below we will, in this work, focus on soft elastic materials that are only slightly rigid. Specifically, we will only consider materials for which $0<G\ll K$. Under these conditions, $a_2\cong0$, $\delta\cong0$, $h(\nu)\cong1$, and the above system of coupled equations are reduced to a single equation. We find:
\begin{equation}\label{eq:gyleq3r}
   P_B=P_e-(4G/3)(1-\nu)+2\gamma/R
\end{equation}
and
\begin{equation}\label{eq:gyleq2r}
   P_B=P_e-\frac{4G}{3}+\frac{2\gamma}{R},
\end{equation}
for a finite and infinite elastic medium, respectively. For reasons also given below, we will here focus on a finite-sized gas bubble embedded in an infinite elastic medium.

One can rewrite Eq. (\ref{eq:gyleq3r}) to define an effective surface tension:
\begin{equation}
\gamma_{eff}\equiv\gamma-\frac{2GR(1-\nu)}{3},
\end{equation}
so that the bubble pressure becomes:
\begin{equation}
P_B=P_e+\frac{2\gamma_{eff}}{R}.
\end{equation}•
Since $2GR(1-\nu)/3\geq0$ for all $R$, we see that the non-zero shear resistance of the elastic medium lowers the effective surface tension acting on the bubble. This holds for either a finite or an infinite elastic medium, but is specific to the type of elastic materials we  consider here (\textit{i.e.} only those which are compressible, and for which $K\gg G$).

\subsection{Generalization of Epstein and Plesset's solution}

Fick's law (Eq. (\ref{eq:fickslaw})) is used in order to obtain the formal growth/dissolution rate expression for a gas bubble embedded in an isotropic elastic medium. Using the Ideal Gas law, we find:
\begin{equation}\label{eq:dndteqfoupietc}
   \frac{dn}{dt}=\frac{4\pi}{3BT}\frac{d}{dt}\left(P_BR^3\right)=4\pi D^*R^2\DCDR.
\end{equation}

Substituting the expression for $P_B$ given by Eq. (\ref{eq:gyleq2r}) into Eq. (\ref{eq:dndteqfoupietc}), gives
\begin{equation}\label{eq:drdtdcdrfunc}
   \frac{dR}{dt}=\frac{3BTD^*}{3P_eR+4\gamma-4GR}\left\{R\DCDR\right\}
\end{equation}
for a bubble in an infinite medium.

In the limit $G\to0$, Eq. (\ref{eq:drdtdcdrfunc}) can be shown to reduce to Epstein and Plesset's rate expressions for $\gamma\geq0$.

\subsection{The $\dcdr$ expressions}

The working rate equations are obtained by replacing $\dcdr$ by the expressions obtained for it from the solution of the diffusion equation. We will consider the expressions for $\dcdr$ that arise both from the diffusion equation, and from its steady-state approximation, the Laplace equation.

It is perhaps not superfluous to point out the distinction between the ``steady-state'' and the ``quasi-static'' approximations, both of which arise in this work. The former entails setting $\partial c/\partial t=0$ in the diffusion equation, which removes its explicit time-dependence.  The time-dependence is then carried implicitly by constants ($r$-independent parameters) determined from the time-dependent boundary conditions (see Eq. (\ref{eq:gralsollapeq})). The quasi-static approximation, on the other hand, involves assuming an instantaneous re-equilibration of the solute distribution in the system, following each incremental gas transfer between the bubble, the surrounding medium, and the WSR. The quasi-static approximation is applicable both to the diffusion equation and to its steady-state approximation (the Laplace equation).

By solving the diffusion equation, Epstein and Plesset found:
\begin{equation}\label{eq:dcdrd2pfunc}
   \DCDR=\frac{P_{WSR}-P_B(t)}{K_H}\left(\frac{1}{R(t)}+\frac{1}{\sqrt{\pi D^*t}}\right),
\end{equation}
or
%
\begin{eqnarray}
   \DCDR&=&(c_{WSR}-c_B(t))\left(\frac{1}{R(t)}+\frac{1}{\sqrt{\pi D^*t}}\right)\nonumber\\
        &=&(fc_{sat}-c_B(t))\left(\frac{1}{R(t)}+\frac{1}{\sqrt{\pi D^*t}}\right).\label{eq:dcdrd2ffunc}
\end{eqnarray}
%
In going from Eq. (\ref{eq:dcdrd2pfunc}) to Eq. (\ref{eq:dcdrd2ffunc}), we used $c_{WSR}\equiv fc_{sat}=fP_e/K_H$, where $c_{sat}$ is the dissolved solute concentration in the well-stirred region of the medium at equilibrium, and $f$ is the relative solute concentration to its equilibrium value (\textit{aka} the ``supersaturation ratio'') in the well-stirred region.

Epstein and Plesset noticed that the $1/\sqrt{\pi D^*t}$ term varied more rapidly with time than $1/R(t)$, and they consequently neglected it in order to derive an approximate analytic expression for the time evolution of the bubble radius. (Their work preceded the computer era so that the numerical procedures that we take for granted were then not an option).

An alternate route to these analytic expressions for the concentration gradient is to solve the Laplace equation:
\begin{equation}\label{eq:lapeq}
   \nabla^2c(r,t)=0
\end{equation}
subject to the boundary conditions:
\begin{equation}\label{eq:cwsrcbdefs}
   c(R_{sys}=\infty)=c_{WSR};\quad c(R,t)=c_B(t).
\end{equation}

The solution of the Laplace equation under spherical symmetry has the general form 
\begin{equation}\label{eq:gralsollapeq}
   c(r,t)=A(t)+\frac{B(t)}{r},\qquad r\geq R(t),
\end{equation}
where $A(t)$ and $B(t)$ are constants with respect to $r$, that are re-evaluated at each time $t$, from the time-dependent boundary conditions. This is an example of the quasi-static approximation, applied to the Laplace equation.

From Eqs. (\ref{eq:cwsrcbdefs}) and (\ref{eq:gralsollapeq}), the concentration gradient at the bubble surface, obtained from the solution of the Laplace equation, is readily found to be
\begin{equation}\label{eq:dcdrl2ffunc}
   \dcdr=(fc_{sat}-c_B(t))/R(t).
\end{equation}
Notice that Eq. (\ref{eq:dcdrd2ffunc}) reduces to Eq. (\ref{eq:dcdrl2ffunc}), after dropping the second term in Eq. (\ref{eq:dcdrd2ffunc}).

\section{Integration of the rate equations}

The rate equation obtained on combining Eqs. (\ref{eq:drdtdcdrfunc}) and (\ref{eq:dcdrd2ffunc}) involves a numerical instability due to the infinite slope at $t=0$. This problem can be eliminated by transforming to $t^{1/2}$ as the time variable. It is also more convenient to work with the equations in dimensionless form. 

Therefore we define dimensionless (or reduced) variables for time, and for the bubble radius as:
\begin{equation}\label{eq:xandrhodef}
   x^2\equiv\left(\frac{2D^*BT}{K_HR_0^2}\right)t,\qquad\rho\equiv\frac{R}{R_0}.
\end{equation}
In Eq. (\ref{eq:xandrhodef}) $R_0$ is the initial bubble radius, and to keep the notation simple, the time-dependencies of $x^2$, $R$, and $\rho$ are not explicitly written but are to be understood.

The semi-regularized dimensionless rate equations are obtained by combining equations (\ref{eq:drdtdcdrfunc}) with (\ref{eq:dcdrd2ffunc}), and (\ref{eq:drdtdcdrfunc}) with (\ref{eq:dcdrl2ffunc}). The results are:
\begin{equation}\label{eq:drhodxd2}
   \frac{d\rho}{dx}=\frac{f-1+\alpha-3\beta/\rho}{\rho(1-\alpha)+2\beta}(x+\lambda\rho),
\end{equation}
and
\begin{equation}\label{eq:drhodxl2}
   \frac{d\rho}{dx}=\left(\frac{f-1+\alpha-3\beta/\rho}{\rho(1-\alpha)+2\beta}\right)x,
\end{equation}
for the diffusion equation-based $\dcdr$, and the Laplace equation-based $\dcdr$, respectively. Here we have defined  
\begin{subequations}
\begin{equation}
   \alpha\equiv4G/(3P_e),\quad\beta\equiv2\gamma/(3R_0P_e),
\end{equation}
and
\begin{equation}
   \lambda\equiv\sqrt{2BT/(K_H\pi)},
\end{equation}
\end{subequations}
as dimensionless constants that are measures of the shear modulus, the surface tension, and the square root of the gas solubility, respectively.

Equations (\ref{eq:drhodxd2}) and (\ref{eq:drhodxl2}), as written, are not fully regularized since they may become singular as $\rho\to0$ (the point at which the bubble dissolves). While this is irrelevant with respect to the Laplace-based Eq. (\ref{eq:drhodxl2}), which can be integrated analytically (below), it does create a problem for dealing with Eq. (\ref{eq:drhodxd2}), which can only be integrated numerically. Therefore, for this numerical integration we adopted a modified Runge-Kutta integration scheme, wherein the behavior of $\rho$ is monitored at the intermediate steps of the integration procedure. The modification involved stopping the calculation and returning the values $\rho=0$ and $\tau=\tau_d$ as the intermediate values of $\rho$ and $\tau$, whenever negative $\rho$ values were encountered. The validity of the method was confirmed by checking it against the analytical solution for the Laplace-based equation (below). The method yielded results whose relative errors oscillated around the errors expected for the traditional 4$^{\textrm{th}}$ order Runge-Kutta integration scheme: $O(10^{-4})-O(10^{-5})$.

As indicated above, Eq. (\ref{eq:drhodxl2}) can be integrated analytically, and the final expression obtained is well-behaved. Using a dimensionless time
\begin{equation}
   \tau\equiv\frac{2D^*BT}{K_HR_0^2}t,
\end{equation}
($x^2=\tau$) the result is:
\begin{eqnarray}
   \tau&=&\frac{1-\alpha}{1-f-\alpha}\left(1-\rho^2\right)
       -\frac{2\beta\left(2f+1-\alpha\right)}{\left(1-f-\alpha\right)^2}\left(1-\rho\right)\nonumber\\
       & &+\frac{6\beta^2\left(2f+1-\alpha\right)}{\left(1-f-\alpha\right)^3}
       \ln\left(\frac{(1-f-\alpha)+3\beta}{\left(1-f-\alpha\right)\rho+3\beta}\right).\label{eq:tausoll2}
\end{eqnarray}
The dissolution time $\tau_d$ is found from:
\begin{equation}
   \tau_d=\tau(\rho=0).
\end{equation}
%

\section{Results}

We first identify some actual materials for which $G\ll K$, some of which are relevant to our interest in modeling gas bubble dynamics in soft extravascular tissue. These include solutions of gelatin in water, and a number of soft, largely aqueous tissues in humans and animals. The shear modulus for gelatin solutions can be made to have a variety of different values by adjusting the concentration of the polymeric gelatin solute, the temperature, the pH, and the concentration of any other solutes that may be present \cite{bib:nijenhuis1997,bib:shanhui1993}. Partly for this reason, aqueous gelatin solutions have been used extensively to model the influence of shear resistance effects in volcanology \cite{bib:giuseppe2009,bib:kavanagh2013}, and in studies related to Decompression Sickness \cite{bib:yount1982}.

In Table \ref{tab:softmatG}, we list values for the shear modulus for gelatin solutions under different conditions, and for a variety human and animal soft tissue. This list is by no means exhaustive, but it illustrates the approximate magnitudes of the reported $G$ values, and some materials to which our expressions would be applicable. The entries at the bottom of the table for limb, muscle and cartilage are probably the most relevant to our interest in modelling gas bubbles responsible for joint pain and for musculoskeletal Decompression Sickness.
\begin{table}[htb!]
\centering
\begin{tabular}{||c|c|c||}
\hline\hline\hline
\textbf{Material} & $\boldsymbol{G}$\textbf{(atm)} & \textbf{Ref.}\\\hline\hline\hline
Gelatin solution & $0.083 - 0.434$ & \citenum{bib:nijenhuis1997} \\\hline
Gelatin solution & $0.0002 - 0.0004$ & \citenum{bib:shanhui1993} \\\hline
Gelatin/Agar  & 0.07 & \citenum{bib:glozman2010} \\\hline\hline\hline
Neural retina & $\sim9.87\times10^{-4}$ & \citenum{bib:forgacs1998} \\\hline
Liver & $0.001-0.003$ & \citenum{bib:chen1996} \\\hline
Liver (Bovine) & $0.10$ & \citenum{bib:glozman2010} \\\hline
Liver & $0.001$ & \citenum{bib:forgacs1998} \\\hline
Heart & $0.001$ & \citenum{bib:forgacs1998} \\\hline\hline\hline
Fat (Porcine) & $0.46$ & \citenum{bib:glozman2010} \\\hline
Breast (Turkey) & $0.10$ & \citenum{bib:glozman2010} \\\hline
Limb & $0.01$ & \citenum{bib:forgacs1998} \\\hline
Muscle & $0.005-0.010$ & \citenum{bib:chen1996} \\\hline
Articular Cartilage & $0.33-5.26$ & \citenum{bib:toyras2001} \\\hline
Knee Cartilage & $2.0-4.0$ & \citenum{bib:wong2008} \\\hline
\hline
\end{tabular}
\caption{Shear modulus for different soft materials for which $G\ll K$.}\label{tab:softmatG}
\end{table}

The large ranges provided above for some of the entries reflect different ways in which the shear resistance was measured, the specific tissue used, and the condition (degree of stress and strain) of the sample at the time of measurement.

The bulk modulus of dilute gelatin solutions will be dominated by the bulk modulus of water, which is known to be   $\sim2.14\times10^{4}$ atm, and actual measurements suggest that the shear and bulk moduli of such gelatin solutions differ by three orders of magnitude (Poisson ratio of 0.4996) \cite{bib:lesperance2013}. Also, the compressibility modulus of soft tissues is usually several orders of magnitude greater than their shear modulus (see Ref. \citenum{bib:sarvazyan1998} and references therein). Consequently, for both dilute gelatin solutions, and for the soft tissues listed in Table \ref{tab:softmatG}, the condition $G \ll K$ is fully satisfied.

The parameter values chosen for our calculations were further constrained by two considerations. 

First, the requirement that the bubble pressure be non-negative requires that $G\leq3P_e/4$ (see Eq. (\ref{eq:gyleq2r})). This requirement stems from the fact that gas bubbles with negative pressures, embedded in a medium with a positive pressure, are unstable on a thermodynamic time scale. Since $P_e$ is here 1 atm, this produces the constraint $G\leq0.75$ atm in this work. This requirement is satisfied by many soft elastic materials, including most of those listed in Table \ref{tab:softmatG}, or by any other material for which $G\ll K$. For materials that don't satisfy the latter requirement, but whose bubble pressure is non-negative (\textit{e.g.} molten magma), the system can be studied by numerically integrating Eq. (\ref{eq:dndteqfoupietc}), and iteratively solving the system of coupled equations (\ref{eq:gpbcoupled}) at each time step.

Second, it is known that as the thickness of the elastic material becomes reduced, the linear response approximation, which we assumed to hold for the relation between the stress and strain tensor components \cite{bib:goldman2009} loses accuracy \cite{bib:stoker1968,bib:ogden1997,bib:fuogden2001}. Quadratic and possibly higher-order terms must then be included \cite{bib:adkins1964}. Consequently, we will restrict our calculations to a finite-sized bubble surrounded by an infinite elastic shell.

Our results are given graphically in Figs \ref{fig:rhovstaudifflap}-\ref{fig:rvstalphaeff}, for which the fixed parameter values were: $T=298.15\ K$, $P_e=1$ atm, $D^*=2900\ \mu^2/$sec, $\gamma=0.7\ \mu\cdot$atm (70 dynes/cm), and $K_H = 1614$ l$\cdot$atm/mol. The values for $D^*$, $\gamma$, and $K_H$, correspond to the diffusion constant of air in water (for the purposes of this paper, we ignore the shift in the diffusion constant due to shear resistance, since any effect due to this shift would not change the overall behavior of our system), the surface tension of water, and the reciprocal of the solubility of air in water at 1 atm, respectively. We use these values to illustrate the general form of our solutions for soft elastic materials of the kind shown in Table \ref{tab:softmatG}.

For purposes of checking and benchmarking, additional numerical values of dissolving times are provided in greater detail in tables, in the separate section \textbf{``Supplementary Information''} \cite{bib:suppinfo}.

\subsection{The effects on bubble growth or dissolution of shear modulus, surface tension, initial bubble radius, and external solute concentration.}

%
\begin{figure}[b!]
\centering
\includegraphics[width=0.45\textwidth]{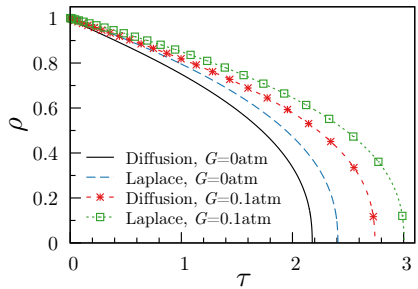}
\caption{Reduced bubble radius as a function of reduced time for a dissolving bubble. The results are for the diffusion and Laplace equations, for fluids with $G\geq0$. The initial bubble radius was the same ($R_0=10\mu$) for all four plots. The plots were obtained using Eq. (\ref{eq:drhodxd2}) and Eq. (\ref{eq:tausoll2}), for the Diffusion and Laplace equations, respectively. The supersaturation ratio ``$f$'' was here set equal to 0.75 for all the plots.}\label{fig:rhovstaudifflap}
\end{figure}

In Fig. \ref{fig:rhovstaudifflap} we show the radius \textit{vs} time dissolution plots obtained by using either the diffusion or the Laplace equations, and the effect of a non-zero value of the shear modulus, with both equations. Clearly, using the diffusion equation, as opposed to its steady-state approximation (the Laplace equation) produces a more rapid dissolution, for a given $G$. This is readily understood from Equations (\ref{eq:dcdrd2ffunc}) and (\ref{eq:dcdrl2ffunc}), from which it is seen that the steady-state approximation, by omitting the $1/\sqrt{\pi D^*t}$ term, results (at all times, and particularly short times) in an insufficiently negative value of the surface gradient term $\dcdr$. The latter controls the dissolution rate through Eqs. (\ref{eq:dndteqfoupietc}) and (\ref{eq:drdtdcdrfunc}).

It is also evident from Fig. \ref{fig:rhovstaudifflap}, that a non-zero shear modulus reduces the rate of dissolution (for either equation), and this is also easily understood. From Eqs. (\ref{eq:henryslawarb}) and (\ref{eq:gyleq2r}), $c_B(G>0)<c_B(G=0)$, so that for a dissolving bubble, for which $\dcdr<0$, $\dcdr$ for $G>0$, is less negative than $\dcdr$ for $G=0$. Consequently (from Eq. (\ref{eq:drdtdcdrfunc})), $dR/dt$ is less negative for $G>0$, so that we get a slower dissolution rate for $G>0$, and this remains true whether one is using the Laplace equation or the diffusion equation.

\begin{figure}[tb!]
\centering
\includegraphics[width=0.45\textwidth]{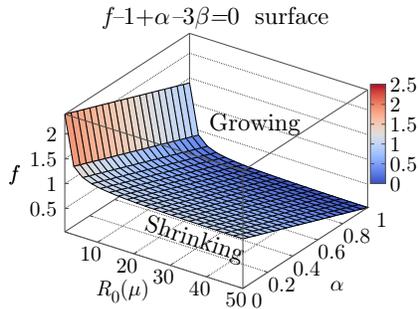}
\caption{Dividing surface for bubble growth and dissolution. The surface is the solution of the equation: $(f-1)+\alpha-3\beta=0$, in terms of the variables $f$, $R_0$, and $\alpha$. All bubbles of initial radius $R_0$ (shown here in microns) lying above the surface will grow; those located below the surface will dissolve.}\label{fig:fvsroalphasurf}
\end{figure}

In Fig. \ref{fig:fvsroalphasurf} we illustrate the effect of the variables ($f,G,R_0$) on bubble growth and dissolution. The surface shown was obtained by setting the numerator in either Eqs. (\ref{eq:drhodxd2}) or (\ref{eq:drhodxl2}) to zero, setting $\rho=1$, and solving the resultant equation for ($f,G,R_0$). A bubble will grow or shrink depending on the sign of $(f-1)+\alpha-3\beta$. As shown (and as expected), bubble dissolution is favored by a small initial radius, a small $G$, and a small $f$, while bubble growth is favored by a large initial radius, a large $G$, and a large $f$. 

\begin{figure}[ht!]
\centering
\includegraphics[width=0.45\textwidth]{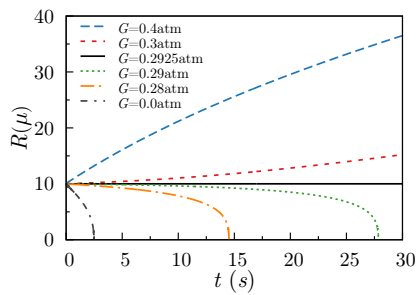}
\caption{The effect of the shear modulus on the evolution of a $10\mu$ bubble embedded in an under-saturated medium ($f=0.75$). The plots were obtained from Eq. (\ref{eq:drhodxd2}). They can be taken to represent the form of the predicted dynamics of a bubble embedded in gelatin solutions, and/or in most of the soft tissues listed in Table \ref{tab:softmatG}.}\label{fig:rvstalphaeff}
\end{figure}

In Figure \ref{fig:rvstalphaeff} we show that a bubble whose elastic diffusion shell is embedded in an under-saturated medium will grow, if the shear modulus of the medium is sufficiently large. The exact value of $G$ for which the transition from dissolution to growth occurs is given by the solution of $(f-1)+\alpha-3\beta=0$. For the parameters used in Fig \ref{fig:rvstalphaeff}, this occurs at $G=0.2925$ atm. For this value of $G$, $f=0.75$, and the other parameter values given previously, a $10\mu$ bubble will be metastable, and may persist at that radius for a significant period of time (Fig. \ref{fig:rvstalphaeff}, black solid line). To the best of our knowledge, this is the first time that such behavior --- bubble growth or meta-stability in an under-saturated medium --- has been theoretically proposed for bubbles surrounded by a homogeneous diffusive medium. Stable bubbles in under-saturated media have been experimentally observed at the liquid-solid interface of some systems (see for example Ref. \citenum{bib:seddonlohse2011}). However, to the best or our knowledge, the meta-stability of bubbles in a homogeneous diffusive medium, which is \textbf{\textit{under-saturated}} with respect to the dissolved gas in the bubble, has not been theoretically proposed elsewhere. The physical basis for this unusual behavior is the negative effect on the gas bubble pressure that arises for the parameter values: $0<G<3P_e/4$; $P_e\approx O(1\ \mathrm{atm})$ (see Eq. (\ref{eq:gyleq2r}), and Ref. \citenum{bib:goldman2009}).

It is worth noting that the growth illustrated in Fig. \ref{fig:rvstalphaeff} is strictly valid only for short times and for a bubble embedded in an infinite elastic medium. For real physical systems and long times, the results we show are approximate limiting values, and the dynamics will deviate somewhat from what is shown here. This is because in a real system, the elastic medium is not infinite, and after the bubble has grown sufficiently, non-linear effects will start to make themselves manifest.

\subsection{Approximate asymptotic growth law for large bubbles}

For bubbles sufficiently large so that the surface tension term can be neglected, $\beta$ can be approximated by zero in Eq. (\ref{eq:drhodxl2}), so that the bubble's radius grows or shrinks, approximately,  according to:
\begin{equation}\label{eq:rhosoll2bzero}
   \rho=\sqrt{1+\left(f_{\!\textit{eff}}-1\right)\tau},
\end{equation}
\begin{equation}
   f_{\!\textit{eff}}=f/(1-\alpha).
\end{equation}

From Eq. (\ref{eq:rhosoll2bzero}), we see that the sign of ($f_{\!\textit{eff}}-1$) determines whether a large bubble in an elastic medium contracts or expands. It will expand if this function is positive; otherwise,  it will contract. Also, it is seen from these equations, that the shear resistance in the elastic medium can be thought of as increasing the value of the effective dissolved solute  concentration in the well-stirred region from $fc_{sat}$ to $f_{\!\textit{eff}}\,\cdot c_{sat}$.

\section{Relation to previous work on visco-elastic materials}

There exists a body of previous work on the elastic effects of a medium on the dynamics of an embedded gas bubble, but significant differences exist on exactly what was meant by the term ``elastic effects''. A particular focus in the past was the dynamics of a bubble in a foam. These studies on foams also differed from one another, depending on the nature of the elastic shell surrounding the bubble \cite{bib:ruckenstein1970,bib:fyrillas2000,bib:kloek2001}; or by requiring the medium to be an infinite viscous liquid \cite{bib:kloek2001,bib:street1968}, or an elastic-plastic material \cite{bib:kunkle1982}. Furthermore, in some cases, the mathematical form of the growth law for the bubble was imposed as an \textit{a priori} assumption \cite{bib:ruckenstein1970}. Also, in some of the previous work it was assumed that the bubble pressure is affected by an \textit{ad-hoc} polynomial term \cite{bib:terrones2003}, or by postulated phenomenological rules \cite{bib:kunkle1982}.

In this work we used a functional form for the bubble pressure which arises entirely from the general theory of elasticity, in the limit of the linear response regime \cite{bib:landau1989}. In addition, all the previous work of which we are aware has ignored the effect of the medium's compression on the diffusion of the dissolved gas within it. We formally included this effect by using Aifantis' fundamental diffusion equation for elastic solids \cite{bib:aifantis1978}.

Despite our somewhat more fundamental approach, and significant differences in the nature of the elastic materials considered, it is noteworthy that we obtained qualitatively similar results for the effect of elasticity of the medium on the bubble dynamics --- for example, a reduction of the bubble's dissolution rate --- as was reported in some of the earlier work (\textit{e.g.}, compare Fig. \ref{fig:rhovstaudifflap} (above), with Ref. \citenum{bib:kloek2001}).

\section{Summary}

By extending the diffusion equation to allow for elastic effects, and using a recently derived Generalized Young-Laplace equation to account for the effect of shear resistance on the pressure in an embedded gas bubble, we derived a generalization for the bubble's growth/dissolution rate equations that is applicable to gas bubbles embedded in soft elastic materials with properties resembling those of many soft tissues in the body. It was shown that if the shear modulus is sufficiently large (but less than $3P_e/4$, for $P_e\approx O$(1 atm)), a gas bubble may be meta-stable and remain at a constant size for some time, or it may grow, in a medium that is under-saturated with respect to the dissolved gaseous solute. To the best of our knowledge, this is the first report of the possibility of gas bubble meta-stability or growth in a homogeneous \textbf{\textit{under-saturated}} medium. Also, this indicates that gas bubbles in extravascular soft tissue in the body arising from decompression will tend to persist for longer times than they would if they were suspended in a simple (non-elastic) liquid containing the same dissolved gas partial pressure. This has implications for the duration of symptoms of ``the bends''.

\section*{Acknowledgements}
   We are grateful to the Natural Sciences and Engineering Research Council of Canada (NSERC) for financial support in the form of a Discovery Grant (6831-2011) to one of us (SG). We also thank Simon Mitchell for his very helpful comments on arterial \textit{vs} extra-vascular forms of Decompression Sickness.

\end{document}


\title{{\LARGE Supplementary Information:}\\{\large ``Gas bubble dynamics in soft materials''}}
\author{J. M. Solano-Altamirano}
\email{jmsolanoalt@gmail.com}
\author{John D. Malcolm}
\email{malcolmj@uoguelph.ca}
\author{Saul Goldman}
\email{sgoldman@uoguelph.ca}
\affiliation{Dept. of Chemistry, the Guelph-Waterloo Centre for Graduate Work in Chemistry and the Guelph-Waterloo Physics Institute, University of Guelph, Guelph, Ontario, N1G 2W1, Canada}

\date{\today}                                           

\begin{abstract}
\textbf{Supp. Info. Abstract: } Here we provide tables of numerical values of dissolving times for a bubble embedded in an soft elastic medium. These values are intended to serve as benchmarks for those readers who may want to check their work against ours.
\end{abstract}

\maketitle

\section{$(\partial c/\partial r)_R$ from the Diffusion equation}

The dissolving time for a bubble embedded in an elastic medium is found by numerically solving the differential equation
\begin{equation}\label{eq:tddiff}
   \frac{dR}{ds}=\frac{6BTD(P_{WSR}-P_e+4G/3-2\gamma/R)}{(3P_eR+4\gamma-4GR)K_H}\left\{s+\frac{R}{\sqrt{\pi D}}\right\},
\end{equation}
which is obtained by combining Eqs. (14), (16), and (17) from the main article, together with the change of variable
\begin{equation}
   t=s^2.
\end{equation}
This change of variable eliminates the singularity at $t=0$ (see Section III of the main article).

A tabulated series of dissolving times for different combinations of the shear modulus $G$, and initial radius $R_0$ is given in Table \ref{tab:tddiff}.

\begin{table}[ht!]
\begin{center}
\begin{tabular}{||c|c|c|c|c||}
\hline
\hline
 & \multicolumn{4}{c||}{$t_d(R_0,G)\times\textrm{sec}^{-1}$}\\
 & \multicolumn{4}{c||}{$\big((\partial c/\partial r)_R\big.$ from the Diffusion equation.$\big.\big)$}\\
\hline
$R_0\times\mu^{-1}$ & $G=0.0$atm & $G=0.1$atm  & $G=0.2$atm  & $G=0.3$atm  \\
\hline
\hline
5 & 0.4741 & 0.5365 & 0.6483 & 0.9195 \\
10 & 2.477 & 3.119 & 4.808 & $\infty$ \\
15 & 6.337 & 8.546 & 16.38 & $\infty$ \\
20 & 12.16 & 17.21 & 40.19 & $\infty$ \\
25 & 20.01 & 29.33 & 82.27 & $\infty$ \\
30 &  29.9 & 45.04 & 150.3 & $\infty$ \\
\hline
\hline
\end{tabular}
\caption{Dissolving times obtained numerically from Eq. (\ref{eq:tddiff}). Here we have used $T=298.15\ K$, $P_e=1$ atm, $P_{WSR}=0.75$ atm, $D=2900\ \mu^2/$sec, $\gamma=0.7\ \mu\cdot$atm (70 dynes/cm), $B=0.082057$ atm$\cdot$l$\cdot$mol$^{-1}\cdot K^{-1}$, and $K_H = 1614$ atm$\cdot$l$\cdot$mol$^{-1}$.}\label{tab:tddiff}
\end{center}
\end{table}

\section{$(\partial c/\partial r)_R$ from the Laplace equation}

The time evolution of a bubble embedded in a soft elastic material, using $(\partial c/\partial r)_R$ obtained from the Laplace equation (Eq. (19) in the main paper), is given by
\begin{eqnarray}
   t&=&\frac{1-\alpha}{2Dd(1-f-\alpha)}(R_0^2-R^2)-\frac{2\gamma(2f+1-\alpha)}{3Dd(1-f-\alpha)^2P_e}(R_0-R)\nonumber\\
    & &+\frac{4\gamma^2(2f+1-\alpha)}{3Dd(1-f-\alpha)^3P_e^2}
         \ln\left(\frac{(1-f-\alpha)R_0P_e+2\gamma}{(1-f-\alpha)RP_e+2\gamma}\right),\label{eq:tevlap}
\end{eqnarray}
where 
\begin{equation}
   d\equiv\frac{BT}{K_H},\qquad f\equiv\frac{P_{WSR}}{P_e},\qquad\textrm{ and }\qquad\alpha\equiv\frac{4G}{3P_e}
\end{equation}
Equation (\ref{eq:tevlap}) has physical units and is equivalent to Eq. (28) of the main paper.

The dissolving times shown in Table \ref{tab:tdlap} were obtained from:
\begin{eqnarray}\label{eq:tdlap}
   t_d&=&\frac{1-\alpha}{2Dd(1-f-\alpha)}R_0^2-\frac{2\gamma(2f+1-\alpha)}{3Dd(1-f-\alpha)^2P_e}R_0\nonumber\\
    & &+\frac{4\gamma^2(2f+1-\alpha)}{3Dd(1-f-\alpha)^3P_e^2}
         \ln\left(\frac{(1-f-\alpha)R_0P_e+2\gamma}{2\gamma}\right),
\end{eqnarray}
which was obtained from Eq (3) with R set equal to 0.

\begin{table}[hp!]
\begin{center}
\begin{tabular}{||c|c|c|c|c||}
\hline
\hline
 & \multicolumn{4}{c||}{$t_d(R_0,G)\times\textrm{sec}^{-1}$}\\
 & \multicolumn{4}{c||}{$\big((\partial c/\partial r)_R\big.$ from the Laplace equation.$\big.\big)$}\\
\hline
$R_0(\mu)$ & $G=0.0$atm & $G=0.1$atm  & $G=0.2$atm  & $G=0.3$atm  \\
\hline
\hline
5 & 0.5316 & 0.5982 & 0.7172 & 1.004 \\
10 &  2.74 & 3.417 & 5.189 & $\infty$ \\
15 & 6.965 & 9.286 & 17.44 & $\infty$ \\
20 & 13.32 & 18.61 & 42.42 & $\infty$ \\
25 & 21.86 &  31.6 & 86.27 & $\infty$ \\
30 & 32.61 & 48.42 & 156.8 & $\infty$ \\
\hline
\hline
\end{tabular}
\caption{Dissolving times obtained from Eq. (\ref{eq:tdlap}). Here we have used $T=298.15\ K$, $P_e=1$ atm, $P_{WSR}=0.75$ atm, $D=2900\ \mu^2/$sec, $\gamma=0.7\ \mu\cdot$atm (70 dynes/cm), $B=0.082057$ atm$\cdot$l$\cdot$mol$^{-1}\cdot K^{-1}$, and $K_H = 1614$ atm$\cdot$l$\cdot$mol$^{-1}$.}\label{tab:tdlap}
\end{center}
\end{table}